# Regression with an infinite number of observations applied to estimating the parameters of the stable distribution using the empirical characteristic function

J. MARTIN VAN ZYL

*Department of Mathematical Statistics and Actuarial Science, University of the Free State, PO Box 339, Bloemfontein, South Africa*

Abstract: A function of the empirical characteristic function exists for the stable distribution, which leads to a linear regression and can be used to estimate the parameters. Two approaches are often used, one to find optimal values of t, but these points are dependent on the unknown parameters. And using a fixed number of values for t. In this work the results when all points in an interval is used, thus where least squares using an infinite number of observations, is approximated. It was found that this procedure performs good in small samples.

## 1 Introduction

Koutrouvelis (1980) showed that a function of the characteristic function leads to a linear regression equation in the unknown parameters of the stable distribution. He derived a least squares regression type estimation procedure based on the empirical characteristic function to estimate the parameters of the stable distribution. For a given sample, the characteristic function, $\phi(t)$, regression equations are formed based on calculating the empirical characteristic function at points $t_1,...,t_K$, with these points in an interval. In this work the approximate least squares regression solution is investigated where $K \to \infty$ points are used, thus when calculating the sample function at every point t in the interval. Consider the least squares approximation where $\phi(t)$ is calculated at every point in an interval, $[x_0, x_0+d], x_0, d > 0$. Suppose the function is calculated at points, $t_j = x_0 + \frac{d}{n} j, j = 1,...,n$, then the approximate least squares solution based on points where $(d/n) \to 0, n \to \infty$, thus all points in the interval $t \in [x_0, x_0 + d], x_0, d > 0$.

Expressions for the design matrix for the regression using an infinite number of points on an interval can be derived, but for the dependent variable an approximation is involved. It was found that the infinite least squares estimators performs extremely good in very small samples and totally outperforms the Kogon and Williams (1998) procedure for more heavy-tailed cases of the stable distribution. At a sample size of say 30 observations, the mean square error is almost halved. Specifically where the tails of the distribution the index is in the region of 1.5, that this estimation procedure leads to excellent performance in small samples.

The characteristic function $\phi(t)$ of the stable distribution is given by

$$\log \phi(t) = -\sigma^\alpha |t|^\alpha \{1 - i\beta sign(t)\tan(\pi\alpha/2)\} + i\mu t, \ \alpha \neq 1,$$

and $\quad \log \phi(t) = -\sigma^\alpha |t|^\alpha \{1 - i\beta sign(t)\log(|t|)\} + i\mu t, \ \alpha = 1.$

The parameters are the index $\alpha \in (0, 2]$, scale parameter $\sigma > 0$, coefficient of skewness, $\beta \in [-1,1]$ and mode $\mu$. The symmetric case with $\mu = 0, \beta = 0$ will be considered in this work. The notation and review is based on the work of Weron, p915 in the book by (Gentle, Härdle, Mori, eds, 2004). Koutrouvelis (1980) made use of the properties of the characteristic function and using the fact that $|\phi(t)|^2 = \exp(-2\sigma^\alpha |t|^\alpha)$ derived the model

$$\log(-\log(|\phi(t)|^2)) = \log(2\sigma^\alpha) + \alpha \log(|t|), \qquad (1)$$

a simple linear regression model can be formed using $\hat{\phi}_n(t)$, calculated at various values of t:

$$y_k = m + \alpha \omega_k + \varepsilon_k. \qquad (2)$$

Thus with $y_k = (-\log(|\hat{\phi}(t_k)|^2))$, $\omega_k = \log(|t_k|)$, $m = \log(2\sigma^\alpha)$ and $\varepsilon_k$ an error term. The characteristic function is estimated for a given value of t, for a sample of size n i.i.d. observations $x_1,...,x_n$, as $\hat{\phi}_n(t) = \frac{1}{n}\sum_{j=1}^{n} e^{itx_j}$.

A problem encountered is that the moments and thus optimal points with respect to heteroscedasticity depends on the unknown parameters to be estimated. Koutrouvelis (1980) suggested using $t_k = \pi k/25, k=1,...,K$, and optimal values of K was suggested for various sample sizes and $\alpha's$. An expression for the covariance $cov(|\hat{\phi}_n(t_j)|^2,|\hat{\phi}_n(t_k)|^2)$ and thus the variance of $|\hat{\phi}_n(t_j)|^2$ is given by Koutrouvelis (1980). Since the expressions depends on the unknown parameters, and thus also $Var(\log(-\log(|\hat{\phi}_n(t_j)|^2)))$, weighted regression is problematic.

Two approaches are used to estimate the parameters using the sample or empirical characteristic function. One approach is that there are optimal points, $t_1,...,t_K$, where the points and also K is a function of the unknown parameters and especially the index $\alpha$ of the stable distribution. One such a procedure is the method of Koutrouvelis (1980) which performs excellent when using the optimal number of points for the specific index, but it is very biased when choosing K incorrectly. The optimal points are $t_1,...,t_K, t_k = \pi k/25, k=1,...,K$, with different K's for different values of the unknown index $\alpha$ of the stable distribution and different values also for different sample sizes.

In the other approach a fixed number of points is chosen for any values of the parameters and all sample sizes. Kogon and Williams (1998) suggested using a fixed 10 points [0.1, 0.2, …,1.0], and these points yields good results overall.

## 2 The approximate regression equation based on an infinite sample

The general approximate equations will be derived for the model with *log(t)* as independent variable and it is shown in (2.2) that the estimation equations for a polynomial approximation can easily be derived using similar principles.

### 2.1 The simple linear regression model with independent variable log(t)

Consider the model $E(y|t) = \beta_0 + \beta_1 \log(t)$, $t \in [x_0, x_0 + d]$, $x_0, d > 0$, with $y = \phi(t)$. Let $S(\beta_0, \beta_1)$ denote the sum of squares to be minimized where $y_j = \beta_0 + \beta_1 \log(x_0 + \frac{d}{K} j)$, and let $t_j = x_0 + \frac{d}{K} j$. A sample of size n i.i.d. observations, $y_1,..., y_n$, are available and

$$\hat{\phi}_n(t_j) = \frac{1}{n}\sum_{i=1}^{n} e^{it_j x_i} \text{ with } z_j = \log(-\log(|\hat{\phi}(t_j)|^2)).$$

$$S(\beta_0, \beta_1) = \lim_{K \to \infty} \frac{1}{K+1} \sum_{j=0}^{K} (z_j - \beta_0 - \beta_1 \log(x_0 + \frac{d}{K} j))^2 . \qquad (3)$$

$$\frac{\partial S(\beta_0,\beta_1)}{\partial \beta_1}=0 \text{ if } \lim_{K\to\infty}\frac{1}{K+1}\sum_{j=0}^{K}\log(x_0+\tfrac{d}{K}j)(z_j-\beta_0-\beta_1\log(x_0+\tfrac{d}{K}j))=0,$$

$$\frac{\partial S(\beta_0,\beta_1)}{\partial \beta_0}=0 \text{ if } \lim_{K\to\infty}\frac{1}{K+1}\sum_{j=0}^{K}(z_j-\beta_0-\beta_1\log(x_0+\tfrac{d}{K}j))=0.$$

Using the expected value expression of a variable defined on the interval, $[x_0, x_0+d]$, with respect to the uniform density function $p(t)=1/d$, it follows that

$$\lim_{n\to\infty}\frac{1}{K+1}\sum_{j=0}^{K}\log^m(x_0+\tfrac{d}{K}j)=(1/d)\int_{x_o}^{x_0+d}\log^m(t)dt. \quad (4)$$

The following result given in Gradshteyn and Ryzhik (1980, p203), will be needed to derive the estimation equations:

$$\int \log^m(t)dt = t\log^m(t) - m\int \log^{m-1}(t)dt.$$

Using this result, it follows that:

$$x_{1,2}=x_{2,1}=\lim_{n\to\infty}\frac{1}{K+1}\sum_{j=0}^{K}\log(x_0+\tfrac{d}{K}j)=(1/d)\int_{x_o}^{x_0+d}\log(t)dt$$

$$=(1/d)\left[(x_0+d)\log(x_0+d)-x_0\log(x_0)-d\right],$$

$$x_{2,2}=\lim_{n\to\infty}\frac{1}{K+1}\sum_{j=0}^{K}\log^2(x_0+\tfrac{d}{K}j)=(1/d)\int_{x_o}^{x_0+d}\log^2(t)dt$$

$$= (1/d)\left[(x_0 + d)\log^2(x_0 + d) - x_0 \log^2(x_0) - 2dx_{1,2}\right].$$

Consider the term: $\hat{\mu}_m^Y = \lim_{K \to \infty} \frac{1}{K+1} \sum_{j=0}^{K} z_j \log^m(x_0 + \frac{d}{K} j)$.

In the context of the estimation problem stated in (1) this will be:

$$\hat{\mu}_m^Y = \lim_{K \to \infty} \frac{1}{K+1} \sum_{j=0}^{K} \log^m(x_0 + jd/K)\left[\log(-\log(|\hat{\phi}(x_0 + jd/K)|^2))\right].$$

This value can be approximated in practice by choosing a large $K$ and for $m = 0,1$ it follows that

$$\hat{\mu}_0^Y = \frac{1}{K+1} \sum_{j=0}^{K} \log(-\log(|\hat{\phi}(x_0 + jd/K)|^2)), \tag{5}$$

and $\hat{\mu}_1^Y = \frac{1}{K+1} \sum_{j=0}^{K} \log(x_0 + jd/K) \log(-\log(|\hat{\phi}(x_0 + jd/K)|^2))$. (6)

Another option would be numerical integration to estimate $\hat{\mu}_m^Y, m = 0,1$, but the approach based on the average was used in this work. The results above leads to the equation

$$\begin{pmatrix} 1 & x_{1,2} \\ x_{2,1} & x_{2,2} \end{pmatrix} \begin{pmatrix} \beta_0 \\ \beta_1 \end{pmatrix} = \begin{pmatrix} \hat{\mu}_Y^0 \\ \hat{\mu}_Y^1 \end{pmatrix}.$$

Let $X = \begin{pmatrix} 1 & x_{1,2} \\ x_{2,1} & x_{2,2} \end{pmatrix}$, $\mathbf{y} = \begin{pmatrix} \hat{\mu}_Y^0 \\ \hat{\mu}_Y^1 \end{pmatrix}$, $\boldsymbol{\beta} = \begin{pmatrix} \alpha \\ \beta \end{pmatrix}$.

The least squares solution in matrix notation is $\boldsymbol{\beta} = X^{-1}\mathbf{y}$. Let

$S^2 = \dfrac{1}{K}\sum_{j=0}^{K}(y_j - \hat{\alpha} - \hat{\beta}\log(x_0 + jd/K))^2$. The covariance matrix of $\hat{\boldsymbol{\beta}}$, assuming a random

error with mean zero and constant variance, errors independently distributed, is

$$\text{cov}(\hat{\boldsymbol{\beta}}) = S^2 X^{-1} \text{ and } Var(\hat{\beta}_1) = S^2/(x_{2,2} - x_{1,2}^2),$$

where $\hat{\beta}_1 = (\hat{\mu}_Y^1 - x_{1,2}\hat{\mu}_Y^0)/(x_{2,2} - x_{1,2}^2)$

*2.1 The simple linear and polynomial regression models*

In this section the general results using the above ideas will be derived. The simple linear

regression model is $E(y|t) = \beta_0 + \beta_1 t$. The least squares (LS) estimation expression to

minimize with respect to $\beta_0$ and $\beta_1$, where $y_j = \beta_0 + \beta_1(x_0 + \frac{d}{K}j)$:

$$S(\beta_0, \beta_1) = \lim_{K \to \infty} \dfrac{1}{K+1}\sum_{j=0}^{K}(y_j - \beta_0 - \beta_1(x_0 + \tfrac{d}{K}j))^2.$$

$$\frac{\partial S(\beta_0,\beta_1)}{\partial \beta_1}=0 \text{ if } \lim_{K\to\infty}\frac{1}{K+1}\sum_{j=0}^{K}(x_0+\tfrac{d}{K}j)(y_j-\beta_0-\beta_1(x_0+\tfrac{d}{K}j))=0,$$

$$\frac{\partial S(\beta_0,\beta_1)}{\partial \beta_0}=0 \text{ if } \lim_{K\to\infty}\frac{1}{K+1}\sum_{j=0}^{K}(y_j-\beta_0-\beta_1(x_0+\tfrac{d}{K}j))=0.$$

The results can also be derived by using the expected value integral

$$\lim_{K\to\infty}\frac{1}{K+1}\sum_{j=0}^{K}(x_0+\tfrac{d}{K}j)^m = (1/d)\int_{x_0}^{x_0+d}t^m dt$$

$$= \tfrac{1}{d(m+1)}\left[(x_0+d)^{m+1}-x_0^{m+1}\right].$$

$$\lim_{K\to\infty}\frac{1}{K+1}\sum_{j=0}^{K}(x_0+\tfrac{d}{K}j) = x_0+d/2.$$

$$\lim_{n\to\infty}\frac{1}{K+1}\sum_{j=0}^{K}(x_0+\tfrac{d}{K}j)^2 = x_0^2+dx_0+d^2/3.$$

Let $\hat{\mu}_Y = \frac{1}{K+1}\sum_{j=0}^{K}y_j$ and $\hat{\mu}_Y^1 = (1/K(K+1))\sum_{j=0}^{K}jy_{(j)}$, where observations $(x_0,y_0),(x_1,y_1),...,(x_n,y_n)$, with $x_j = x_0+(d/n)j$, $j=0,...,n$ are available.

$$\lim_{K\to\infty}\frac{1}{K+1}\sum_{j=0}^{K}y_j(x_0+\tfrac{d}{K}j) \approx x_0\hat{\mu}_Y + d\hat{\mu}_Y^1.$$

Let $X = \begin{pmatrix} 1 & x_0 + d/2 \\ x_0 + d/2 & x_0^2 + dx_0 + d^2/3 \end{pmatrix}$, $\mathbf{y} = \begin{pmatrix} \hat{\mu}_Y \\ x_0 \hat{\mu}_Y + d\hat{\mu}_Y^1 \end{pmatrix}$, $\boldsymbol{\beta} = \begin{pmatrix} \beta_0 \\ \beta_1 \end{pmatrix}$.

The least squares equation in matrix notation is:

$$X\boldsymbol{\beta} = \mathbf{y} \text{ and } \boldsymbol{\beta} = X^{-1}\mathbf{y}.$$

If $\hat{\mu}_Y, \hat{\mu}_Y^1$ are estimates based on $K$ values of $t$, thus $t_j = x_0 + jd/K$, $j = 1,\ldots,K$, then

$$S^2 = \frac{1}{K}\sum_{j=0}^{K}(y_j - \hat{\beta}_0 - \hat{\beta}_1(x_0 + jd/K))^2.$$ The covariance matrix of $\hat{\boldsymbol{\beta}}$, assuming a random error with mean zero and constant variance, independently distributed, is $\text{cov}(\hat{\boldsymbol{\beta}}) = S^2 X^{-1}$.

The determinant of the information matrix of the design, X is $d^2/12$. If $Var(y|t) = \sigma^2$, and the residuals are independent with at least a finite second moment, then this variance can be used to test hypothesis concerning $\beta_1$.

The results can be generalized to fit a polynomial. Let $\lim_{K \to \infty} \frac{1}{K+1}\sum_{j=0}^{K} y_j(x_0 + \frac{d}{K}j)^m = \mu_Y^m$. To fit a polynomial of degree m, the following equation must be solved:

$$\begin{pmatrix} x_{1,1} & \cdots & x_{1,m+1} \\ \vdots & \ddots & \vdots \\ x_{m+1,1} & \cdots & x_{m+1,m+1} \end{pmatrix} \begin{pmatrix} \beta_0 \\ \beta_1 \\ \vdots \\ \beta_m \end{pmatrix} = \begin{pmatrix} \hat{\mu}_Y^0 \\ \hat{\mu}_Y^1 \\ \vdots \\ \hat{\mu}_Y^m \end{pmatrix},$$

with $x_{i,j} = \frac{1}{d(l+1)}\left[(x_0+d)^{l+1} - x_0^{l+1}\right]$, $l = i+j-2$, $\hat{\mu}_Y^l = \frac{1}{K+1}\sum_{j=0}^{n} y_j(x_0 + jd/K)^l$, $l = 0,...,K$,

for the function calculated at points $(x_0 + j/K, y_j), j = 0,...,K$.

### 3 A simulation study to investigate the performance of the estimators

In this section a simulation study was conduced to investigate the performance of estimation of the symmetric stable distribution, using equation (1) and based on an approximation of the results above where the characteristic function is approximated using points in the interval [0.1, 2]. The variables in the estimating equations which involve the observed sample observations as in (5) and (6) will be approximated. It was found that the method of estimation proposed in this work performs extremely good in small sample of less than say 200 and especially when the index is in the approximate range of between say 1.0 and 1.7.

It should be noted that standardization of the observations was suggested by some authors, but since this work is conducted on standard stable random variables, it was not standardized. The method of standardization can also influence results.

Suppose $K$ points are used to approximate $\mu_0^Y$ and $\mu_1^Y$ where

$$\hat{\mu}_0^Y = \frac{1}{K+1} \sum_{j=0}^{K} \log(-\log(|\hat{\phi}(x_0 + jd/K)|^2))),$$

and $\hat{\mu}_1^Y = \frac{1}{K+1} \sum_{j=0}^{n_A} \log(x_0 + jd/K) \log(-\log(|\hat{\phi}(x_0 + jd/K)|^2))$.

Since this is a simulation study, with many repetitions, $K$ was chosen not too large. For an individual problem it would be best to choose $K$ large. In this study $K = 500$ was used. The results in table 1 are based on 10000 samples of n=100 each from the symmetric stable distribution with parameters $\sigma = 1.0, \alpha = 1.5$. The interval was chosen as [0.1, 2.0] where the values were calculated for the regression. This interval was chosen as it was found by experimenting, to yield better results than the [0.1,1] interval.

| $K$ | Bias | MSE |
|---|---|---|
| 100 | -0.0422 | 0.0265 |
| 300 | -0.0048 | 0.0239 |
| 500 | 0.0049 | 0.0238 |

**Table 1** Influence of the number of points, $K$, on the estimation of $\alpha$, where the function is calculated to form the estimating equations respect to bias and MSE, $\sigma = 1.0, \mu = 0, \beta = 0, \alpha = 1.5$. Each sample size is 100 and m=10000 generated.

In the following tables estimation results based on m=10000 generated samples are given. The simulation study was conducted with sample sizes up to n=200, since at this sample size the Kogon-Williams estimators start outperforming the approximate infinite number of points procedure.

| $\alpha$ | Koutrouvelis (K optimal \|true $\alpha$) | | | Kogon-Williams | | | LS Infinite Approximation | | |
|---|---|---|---|---|---|---|---|---|---|
| | $\hat{\alpha}$ | Bias | MSE | $\hat{\alpha}$ | Bias | MSE | $\hat{\alpha}$ | Bias | MSE |
| 1.9 | 1.9254 | -0.0254 | 0.0300 | 1.9339 | -0.0339 | 0.0277 | 1.7785 | 0.1215 | 0.0552 |
| 1.5 | 1.5718 | -0.0718 | 0.0972 | 1.5941 | -0.0941 | 0.1064 | 1.4661 | 0.0339 | 0.0693 |
| 1.3 | 1.1456 | 0.1544 | 0.0730 | 1.3849 | -0.0849 | 0.1154 | 1.2874 | 0.0126 | 0.0678 |
| 1.1 | 0.9672 | 0.1328 | 0.0583 | 1.1732 | -0.0732 | 0.1055 | 1.1005 | -0.0005 | 0.0614 |
| 0.9 | 0.7784 | 0.1216 | 0.0468 | 0.9517 | -0.0517 | 0.0834 | 0.9057 | -0.0057 | 0.0521 |
| 0.7 | 0.6233 | 0.0767 | 0.0310 | 0.7303 | -0.0303 | 0.0623 | 0.7060 | -0.0060 | 0.0411 |

**Table 2** Comparison of estimation procedures of $\alpha$ with respect to bias and MSE, $\sigma=1.0, \mu=0, \beta=0, n=30$. The Koutrouvelis results are chosen using the true find the sample size using the true parameters.

| $\alpha$ | Koutrouvelis (K optimal \|true $\alpha$) | | | Kogon-Williams | | | LS Infinite Approximation | | |
|---|---|---|---|---|---|---|---|---|---|
| | $\hat{\alpha}$ | Bias | MSE | $\hat{\alpha}$ | Bias | MSE | $\hat{\alpha}$ | Bias | MSE |
| 1.9 | 1.9185 | -0.0185 | 0.0212 | 1.9251 | -0.0251 | 0.0205 | 1.8126 | 0.0874 | 0.0351 |
| 1.5 | 1.5332 | -0.0332 | 0.0524 | 1.5591 | -0.0591 | 0.0669 | 1.4831 | 0.0169 | 0.0433 |
| 1.3 | 1.1767 | 0.1233 | 0.0452 | 1.3554 | -0.0554 | 0.0664 | 1.3024 | -0.0024 | 0.0416 |
| 1.1 | 0.9967 | 0.1033 | 0.0356 | 1.1410 | -0.0410 | 0.0579 | 1.1047 | -0.0047 | 0.0365 |
| 0.9 | 0.9275 | -0.0275 | 0.0710 | 0.9544 | -0.0544 | 0.0915 | 0.8998 | 0.0002 | 0.0552 |
| 0.7 | 0.7188 | -0.0188 | 0.0520 | 0.7326 | -0.0326 | 0.0621 | 0.7065 | -0.0065 | 0.0405 |

**Table 3** Comparison of estimation procedures of $\alpha$ with respect to bias and MSE, $\sigma = 1.0, \mu = 0, \beta = 0, n = 50$. The Koutrouvelis results are chosen using the true find the sample size using the true parameters.

| $\alpha$ | Koutrouvelis (K optimal \|true $\alpha$) | | | Kogon-Williams | | | LS Infinite Approximation | | |
|---|---|---|---|---|---|---|---|---|---|
| | $\hat{\alpha}$ | Bias | MSE | $\hat{\alpha}$ | Bias | MSE | $\hat{\alpha}$ | Bias | MSE |
| 1.9 | 1.9103 | -0.0103 | 0.0129 | 1.9139 | -0.0139 | 0.0134 | 1.8535 | 0.0465 | 0.0189 |
| 1.5 | 1.5189 | -0.0189 | 0.0266 | 1.5325 | -0.0325 | 0.0344 | 1.5016 | -0.0016 | 0.0238 |
| 1.3 | 1.2147 | 0.0853 | 0.0243 | 1.3288 | -0.0288 | 0.0323 | 1.3109 | -0.0109 | 0.0229 |
| 1.1 | 1.0361 | 0.0639 | 0.0183 | 1.1177 | -0.0177 | 0.0262 | 1.1120 | -0.0120 | 0.0199 |
| 0.9 | 0.8477 | 0.0523 | 0.0137 | 0.9141 | -0.0141 | 0.0212 | 0.9107 | -0.0107 | 0.0162 |
| 0.7 | 0.6787 | 0.0213 | 0.0098 | 0.7100 | -0.0100 | 0.0156 | 0.7104 | -0.0104 | 0.0126 |

**Table 4** Comparison of estimation procedures of $\alpha$ with respect to bias and MSE, $\sigma = 1.0, \mu = 0, \beta = 0, n = 100$. The Koutrouvelis results are chosen using the true find the sample size using the true parameters.

The mean square errors are plotted in figure 1.

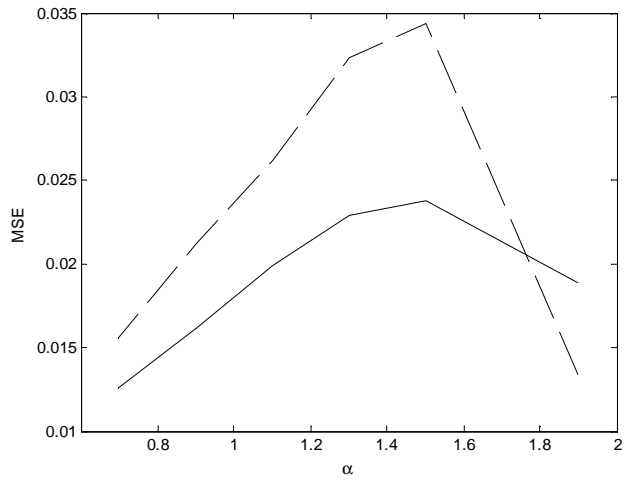

Figure 1. MSE for various values of $\alpha$. Sample size n=100, Kogon-Williams dashed line and infinite approximation solid line.

| $\alpha$ | Koutrouvelis (K optimal \|true $\alpha$) | | | Kogon-Williams | | | LS Infinite Approximation | | |
|---|---|---|---|---|---|---|---|---|---|
| | $\hat{\alpha}$ | Bias | MSE | $\hat{\alpha}$ | Bias | MSE | $\hat{\alpha}$ | Bias | MSE |
| 1.9 | 1.9062 | -0.0062 | 0.0068 | 1.9085 | -0.0085 | 0.0075 | 1.8814 | 0.0186 | 0.0104 |
| 1.5 | 1.5093 | -0.0093 | 0.0133 | 1.5175 | -0.0175 | 0.0174 | 1.5130 | -0.0130 | 0.0137 |
| 1.3 | 1.2463 | 0.0537 | 0.0128 | 1.3131 | -0.0131 | 0.0156 | 1.3147 | -0.0147 | 0.0124 |
| 1.1 | 1.0649 | 0.0351 | 0.0097 | 1.1108 | -0.0108 | 0.0128 | 1.1149 | -0.0149 | 0.0102 |
| 0.9 | 0.8734 | 0.0266 | 0.0076 | 0.9053 | -0.0053 | 0.0097 | 0.9107 | -0.0107 | 0.0080 |
| 0.7 | 0.6919 | 0.0081 | 0.0056 | 0.7044 | -0.0044 | 0.0078 | 0.7078 | -0.0078 | 0.0062 |

**Table 5** Comparison of estimation procedures of $\alpha$ with respect to bias and MSE, $\sigma = 1.0, \mu = 0, \beta = 0, n = 200$. The Koutrouvelis results are chosen using the true find the sample size using the true parameters.

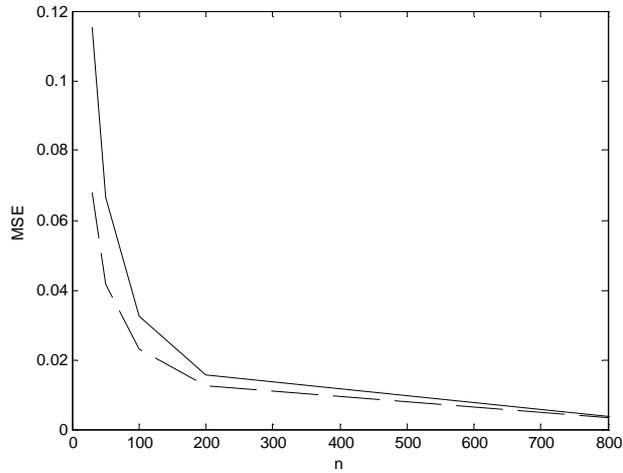

Figure 2. MSE of estimated index for various sample sizes, Kogon-Williams solid line and infinite approximation dashed line.

In tables 6 to 9 results are given for the estimated scale parameter. The performance of the infinite number of points approximation is good with respect to MSE.

| $\sigma$ | Koutrouvelis (K optimal \|true $\alpha$ ) | | | Kogon-Williams | | | LS Infinite Approximation | | |
|---|---|---|---|---|---|---|---|---|---|
| | $\hat{\alpha}$ | Bias | MSE | $\hat{\alpha}$ | Bias | MSE | $\hat{\alpha}$ | Bias | MSE |
| 1.9 | 0.9775 | 0.0225 | 0.0226 | 0.9822 | 0.0178 | 0.0221 | 0.8967 | 0.1033 | 0.0249 |
| 1.5 | 0.9794 | 0.0206 | 0.0390 | 0.9902 | 0.0098 | 0.0414 | 0.9142 | 0.0858 | 0.0320 |
| 1.3 | 0.8431 | 0.1569 | 0.0489 | 0.9840 | 0.0160 | 0.0570 | 0.9221 | 0.0779 | 0.0395 |
| 1.1 | 0.8522 | 0.1478 | 0.0549 | 0.9835 | 0.0165 | 0.0782 | 0.9317 | 0.0683 | 0.0503 |
| 0.9 | 0.8644 | 0.1356 | 0.0749 | 0.9794 | 0.0206 | 0.1164 | 0.9406 | 0.0594 | 0.0750 |
| 0.7 | 0.9123 | 0.0877 | 0.1411 | 0.9750 | 0.0250 | 0.1924 | 0.9532 | 0.0468 | 0.1341 |

**Table 6** Comparison of estimation procedures of $\alpha$ with respect to bias and MSE,

$\sigma =1.0, \mu = 0, \beta = 0, n = 30$. The Koutrouvelis results are chosen using the true find the sample size using the true parameters.

| $\sigma$ | Koutrouvelis (K optimal \|true $\alpha$) | | | Kogon-Williams | | | LS Infinite Approximation | | |
|---|---|---|---|---|---|---|---|---|---|
| | $\hat{\alpha}$ | Bias | MSE | $\hat{\alpha}$ | Bias | MSE | $\hat{\alpha}$ | Bias | MSE |
| 1.9 | 0.9878 | 0.0122 | 0.0140 | 0.9910 | 0.0090 | 0.0137 | 0.9276 | 0.0724 | 0.0158 |
| 1.5 | 0.9825 | 0.0175 | 0.0220 | 0.9958 | 0.0042 | 0.0256 | 0.9478 | 0.0522 | 0.0197 |
| 1.3 | 0.8891 | 0.1109 | 0.0296 | 0.9923 | 0.0077 | 0.0344 | 0.9542 | 0.0458 | 0.0247 |
| 1.1 | 0.8992 | 0.1008 | 0.0344 | 0.9882 | 0.0118 | 0.0501 | 0.9583 | 0.0417 | 0.0335 |
| 0.9 | 0.9594 | 0.0406 | 0.1008 | 0.9787 | 0.0213 | 0.1263 | 0.9349 | 0.0651 | 0.0825 |
| 0.7 | 0.9618 | 0.0382 | 0.1557 | 0.9757 | 0.0243 | 0.1895 | 0.9513 | 0.0487 | 0.1287 |

**Table 7** Comparison of estimation procedures of $\alpha$ with respect to bias and MSE, $\sigma =1.0, \mu = 0, \beta = 0, n = 50$. The Koutrouvelis results are chosen using the true find the sample size using the true parameters.

| $\sigma$ | Koutrouvelis (K optimal \|true $\alpha$) | | | Kogon-Williams | | | LS Infinite Approximation | | |
|---|---|---|---|---|---|---|---|---|---|
| | $\hat{\alpha}$ | Bias | MSE | $\hat{\alpha}$ | Bias | MSE | $\hat{\alpha}$ | Bias | MSE |
| 1.9 | 0.9935 | 0.0065 | 0.0071 | 0.9950 | 0.0050 | 0.0069 | 0.9568 | 0.0432 | 0.0085 |
| 1.5 | 0.9926 | 0.0074 | 0.0116 | 0.9994 | 0.0006 | 0.0134 | 0.9740 | 0.0260 | 0.0111 |
| 1.3 | 0.9350 | 0.0650 | 0.0151 | 0.9986 | 0.0014 | 0.0177 | 0.9797 | 0.0203 | 0.0138 |
| 1.1 | 0.9430 | 0.0570 | 0.0172 | 0.9918 | 0.0082 | 0.0243 | 0.9798 | 0.0202 | 0.0174 |
| 0.9 | 0.9516 | 0.0484 | 0.0228 | 0.9920 | 0.0080 | 0.0364 | 0.9826 | 0.0174 | 0.0251 |
| 0.7 | 0.9735 | 0.0265 | 0.0356 | 0.9933 | 0.0067 | 0.0590 | 0.9870 | 0.0130 | 0.0397 |

**Table 8** Comparison of estimation procedures of $\alpha$ with respect to bias and MSE, $\sigma = 1.0, \mu = 0, \beta = 0, n = 100$. The Koutrouvelis results are chosen using the true find the sample size using the true parameters.

| $\sigma$ | Koutrouvelis (K optimal \|true $\alpha$) | | | Kogon-Williams | | | LS Infinite Approximation | | |
|---|---|---|---|---|---|---|---|---|---|
| | $\hat{\alpha}$ | Bias | MSE | $\hat{\alpha}$ | Bias | MSE | $\hat{\alpha}$ | Bias | MSE |
| 1.9 | 0.9971 | 0.0029 | 0.0036 | 0.9980 | 0.0020 | 0.0036 | 0.9749 | 0.0251 | 0.0048 |
| 1.5 | 0.9963 | 0.0037 | 0.0059 | 1.0003 | -0.0003 | 0.0067 | 0.9881 | 0.0119 | 0.0061 |
| 1.3 | 0.9625 | 0.0375 | 0.0082 | 0.9988 | 0.0012 | 0.0090 | 0.9902 | 0.0098 | 0.0076 |
| 1.1 | 0.9688 | 0.0312 | 0.0093 | 0.9962 | 0.0038 | 0.0124 | 0.9898 | 0.0102 | 0.0092 |
| 0.9 | 0.9770 | 0.0230 | 0.0119 | 0.9963 | 0.0037 | 0.0183 | 0.9920 | 0.0080 | 0.0127 |
| 0.7 | 0.9883 | 0.0117 | 0.0181 | 0.9962 | 0.0038 | 0.0303 | 0.9921 | 0.0079 | 0.0204 |

**Table 9** Comparison of estimation procedures of $\alpha$ with respect to bias and MSE, $\sigma = 1.0, \mu = 0, \beta = 0, n = 200$. The Koutrouvelis results are chosen using the true find the sample size using the true parameters.

*4 Conclusions*

The approximation of the infinite number of observations least squares is not difficult from a computational viewpoint. It was found in this study that this approach leads to good estimation results in small samples for some common values of the index which often occur in practice, where the mean is finite and the variance maybe not.

There is much research which still can be conducted to improve on the ideas in this work, for example the interval and the methods of approximating the estimating equations.